\documentclass[11pt]{article}
\usepackage[a4paper]{geometry}
\usepackage[utf8]{inputenc}
\usepackage{graphicx}
\usepackage{dsfont}
\usepackage{fancyhdr}
\usepackage{booktabs}
\usepackage[colorlinks=true, allcolors=blue,pagebackref=true]{hyperref}
\usepackage{url}
\usepackage{xcolor}
\usepackage{colortbl}
\usepackage{subfig}
\usepackage{mathtools}
\usepackage[T1]{fontenc}
\usepackage{comment}
\usepackage{natbib}
\usepackage{eurosym}
\usepackage{babel}
\usepackage{amsmath}
\allowdisplaybreaks
\usepackage{xcolor}
\usepackage{enumitem}
\usepackage{float}
\usepackage{here}
\usepackage{wrapfig}
\usepackage{lscape}
\usepackage{bbold}
\usepackage{stmaryrd}
\usepackage[export]{adjustbox}
\usepackage{tabularx}
\usepackage{amssymb}
\usepackage{mathtools}
\usepackage{multirow,makecell, framed}
\usepackage[onelanguage, boxruled, vlined]{algorithm2e}
\SetAlFnt{\footnotesize}
\usepackage{amsthm, bm}
\usepackage{lscape}
\usepackage{ulem}
\usepackage{vmargin}
\usepackage{siunitx}

\newcommand{\dint}{\displaystyle\int}
\newcommand{\dsum}{\displaystyle\sum}

\newcommand\cites[1]{\citeauthor{#1}'s \citeyearpar{#1}}

\newtheorem{lemma}{Lemma}

\newtheorem{theorem}{Theorem}
\newtheorem{definition}{Definition}

\theoremstyle{remark}
\newtheorem{remark}{Remark}

\usepackage{authblk}

\renewcommand*{\backref}[1]{}
\renewcommand*{\backrefalt}[4]{}

\setmarginsrb{2cm}{1cm}{2cm}{1cm}{0.5cm}{0.5cm}{0.5cm}{0.5cm}

\title{Investigating new, signature-based, spatial autoregressive models for functional covariates}
\author[,1]{Camille Frévent\thanks{Corresponding author: \texttt{camille.frevent@univ-lille.fr}}}
\affil[1]{Univ. Lille, CHU Lille, ULR 2694 - METRICS: Évaluation des technologies de santé et des pratiques médicales, F-59000 Lille, France}

\date{}

\begin{document}

\maketitle

\hrule
\section*{Abstract}
We developed two new alternatives to signature-based, spatial autoregressive models. In a simulation study, we found that the new models performed at least as well as existing approaches but presented shorter computation times. We then used the new models to analyze the premature mortality rate and the mortality rate for people aged 65 and over. \\[0.2cm]
\textbf{Keywords:} Functional data, Partial least squares regression, Signature, Spatial regression, Tensor \vspace{0.5cm}
\hrule

\section{Introduction}

Advances in sensing technology and greater data storage capacities have increased the volume of continuously recorded data. This change prompted the development of functional data analysis by \cite{ramsaylivre} and the subsequent adaptation of many statistical methods for use with a functional framework. 

Here, we focus on regression models involving a real-valued response variable and functional covariates observed over a time interval $\mathcal{T}$. Conventionally, one assumes that the functional covariate $X$ belongs to $\mathcal{L}^2(\mathcal{T}, \mathbb{R}^p)$, the space of $p$-dimensional square-integrable functions on $\mathcal{T}$, and considers the following linear model \citep{ramsaylivre}:
$$ Y = \dint_{\mathcal{T}} X(t)^\top \beta^*(t)  \ \text{d}t + \varepsilon. $$ 

This model is typically estimated by approximating $X$ and $\beta^*$ as finite combinations of basis functions, which thereby reduces the problem to a classical linear regression on the resulting coefficients. \\

In recent years, signatures have gained popularity in a number of fields, including medicine \citep{moore2019using, morrill2020utilization, falcioni2023path}, character recognition \citep{yang2016rotation, lai2019offline}, and finance \citep{buehler2020generating,adachi2021discrete,inzirillo2024clustering}. \\

Although signatures were initially defined by \cite{chen1957integration, chen1977iterated} for smooth paths and then rediscovered as part of rough path theory \citep{lyons1998differential}, \cite{fermanian2022functional} developed a linear regression model that used the functional covariates' signatures rather than the covariates themselves. \cites{fermanian2022functional} work highlighted three main advantages of the signature-based approach: signatures (i) do not require $X \in \mathcal{L}^2(\mathcal{T}, \mathbb{R}^p)$, (ii) are naturally suited to multivariate functions, and (iii) effectively encode the geometric properties of $X$. \\

In domains in which data naturally involve a spatial dimension (e.g., environmental science), functional data analysis has led to the development of specific methodologies for spatial functional data. In the context of spatial regression, \cite{huang2018spatial} and \cite{ahmed2022quasi} developed a functional, spatial autoregressive model (FSARLM). Assuming that $X \in \mathcal{L}^2(\mathcal{T}, \mathbb{R})$, the FSARLM is defined as 
$$ Y_i = \rho^* \dsum_{j=1}^n w_{ij,n} Y_j + \dint_\mathcal{T} \beta(t)^* X_i(t) \ \text{d}t + \varepsilon_i $$
where $W_n = (w_{ij,n})_{1 \le i,j \le n}$ is a non-stochastic spatial weight matrix, $\rho^*$ is a spatial autoregressive parameter in $[-1,1]$, $\beta(t)^* \in \mathcal{L}^2(\mathcal{T}, \mathbb{R})$, and the disturbances $\varepsilon_i$ are assumed to be independent, identically distributed, and independent of $\{X_i(\cdot), i = 1, \dots, n\}$. \\

\cite{Fre2025} developed two signature-based approaches (penalized signature-based spatial regression (PenSSAR) and a spatial autoregressive model based on signature projections (ProjSSAR)) for use with functional covariates. PenSSAR is based on ridge penalized regression, and ProjSSAR is based on a principal component analysis (PCA). The results of a simulation study showed that PenSSAR performed better than ProjSSAR and the FSARLM. However, the PenSSAR estimation procedure is iterative and may therefore require a long computation time. Recently, \cite{frevent2024multivariate} developed an autoregressive spatial model for functional covariates in the context of a multivariate response variable using signatures, which presents the advantage of not using an iterative algorithm for the estimation. \\
In this article, we first describe our adaptation of \cites{frevent2024multivariate} approach for the case of a univariate response variable (NaivePenSSAR). Furthermore, we have developed a new approach based on a partial least squares (PLS) projection. To distinguish between the two projection-based approaches, we refer to the PCA approach as PCA-ProjSSAR and our new PLS strategy as PLS-ProjSSAR.

The signatures and properties are described in Section \ref{sec:sig}. In Section \ref{sec:methodology}, we describe our adaptation of \cites{frevent2024multivariate} approach for the case of a univariate response variable and our new PLS-ProjSSAR.
In Section \ref{sec:simu}, we describe a simulation study in which the new models were compared with the FSARLM \citep{ahmed2022quasi}, PenSSAR and PCA-ProjSSAR \citep{Fre2025} approaches. Next, the methods were applied to a real dataset, as described in Section \ref{sec:appli}. Lastly, we discuss these various approaches in Section \ref{sec:discussion}.

\section{The signature concept} \label{sec:sig}

Here, we briefly present the signature concept; for a comprehensive description, we refer the reader to \cite{lyons2007differential} and \cite{friz2010multidimensional}. Although signatures were initially developed for smooth paths, we employ them here in the context of functional data and have adopted standard notation for clarity.

\begin{definition}[Function of bounded variation] \phantom{linebreak} \\
Let $X: \mathcal{T} \rightarrow \mathbb{R}^p$. The total variation of $X$ is defined by 
$$
\left\| X\right\|_{TV} = \sup_{(t_0,...,t_k)\in \mathcal{I}} \sum_{i=1}^k \left\| X(t_i) - X(t_{i-1}) \right\|,
$$
where $\left\| .\right\|$ denotes the Euclidean norm on $\mathbb{R}^p$ and $\mathcal{I}$ denotes the set of all finite partitions of $\mathcal{T}$. 
$X$ is said to be of bounded variation if $\left\| X\right\|_{TV} < +\infty$. \\

By $\mathcal{C}(\mathcal{T},\mathbb{R}^p)$, we denote the set of $p$-dimensional continuous functions of bounded variation on $\mathcal{T}$.
\end{definition}

\begin{definition}[The signature of $X \in \mathcal{C}(\mathcal{T},\mathbb{R}^p)$] \phantom{linebreak} \\
Let $\mathcal{X} \in \mathcal{C}(\mathcal{T},\mathbb{R}^p)$. The signature of $\mathcal{X}$ is the following sequence 
$$ Sig(X) = (1, \bm{X}^1, \dots, \bm{X}^d, \dots), $$
where $$ \bm{X}^d = \underset{\substack{t_1 < \dots < t_d \\ t_1, \dots, t_d \in \mathcal{T}}}{\dint \dots \dint} \ \text{d}X(t_1) \otimes \dots \otimes \text{d}X(t_d) \in \left(\mathbb{R}^{p}\right)^{\otimes d}. $$
\end{definition}

\begin{definition}[The signature coefficient \citep{levin2013learning}] \phantom{linebreak} \\
Let $A^*$ be the set of multi-indexes with entries in $\{1, \dots, p \}$. Let $\{e_i, i = 1, \dots, p\}$ be the canonical orthonormal basis of $\mathbb{R}^p$. For any positive integer $d$, the space $(\mathbb{R}^p)^{\otimes d}$ is isomorphic to the free vector space generated by all the words of length $d$ in $A^*$ and $(e_{i_1} \otimes \dots \otimes e_{i_d})_{(i_1, \dots, i_d) \in A^*}$ form a basis of $(\mathbb{R}^p)^{\otimes d}$. Therefore, the signature of $X \in \mathcal{C}(\mathcal{T},\mathbb{R}^p)$ can be rewritten
as
$$ Sig(X) = 1 + \dsum_{d=1}^{\infty} \dsum_{(i_1, \dots, i_d)} \mathcal{S}_{(i_1, \dots, i_d)}(X) e_{i_1} \otimes \dots \otimes e_{i_d},$$
where
$$ \mathcal{S}_{(i_1, \dots, i_d)}(X) = \underset{\substack{t_1 < \dots < t_d \\ t_1, \dots, t_d \in \mathcal{T}}}{\dint \dots \dint} \ \text{d}X^{(i_1)}(t_1) \dots \text{d}X^{(i_d)}(t_d) \in \mathbb{R} $$ is the signature coefficient of order $d$ along $(i_1, \dots, i_d)$ of $X$. \\
Thus, the signature of $X$ is valued in $T((\mathbb{R}^p))$ the tensor algebra space defined as
$$T((\mathbb{R}^p)) = \{ (a_0, \dots, a_d, \dots), \forall d \ge 0, a_d \in (\mathbb{R}^p)^{\otimes d} \} .$$
\end{definition}

\begin{definition}[The truncated signature] \phantom{linebreak} \\
Let $\mathcal{X} \in \mathcal{C}(\mathcal{T},\mathbb{R}^p)$.
The truncated signature of $X$ at order $D \ge 1$ is defined as
$$ Sig^D(X) = (1, \bm{X}^1, \dots, \bm{X}^D). $$
\end{definition}

\begin{definition}[The truncated signature coefficient vector] \phantom{linebreak} \\
Let $\mathcal{X} \in \mathcal{C}(\mathcal{T},\mathbb{R}^p)$. The truncated signature coefficient vector at order $D \ge 1$ of $X$ is defined as the following sequence 
$$
S^D(X)=\left(1,\mathcal{S}_{(1)}(X), \dots, \mathcal{S}_{(p)}(X), \mathcal{S}_{(1,1)}(X), \mathcal{S}_{(1,2)}(X), \dots, \mathcal{S}_{\underbrace{\scriptstyle \left(p, \dots, p \right)}_{D \text{terms}}}(X) \right),
$$
and the truncated shifted-signature coefficient vector at order $D \ge 1$ of $X$ is defined as
$$
\widetilde{S}^D(X)=\left(\mathcal{S}_{(1)}(X), \dots, \mathcal{S}_{(p)}(X), \mathcal{S}_{(1,1)}(X), \mathcal{S}_{(1,2)}(X), \dots, \mathcal{S}_{\underbrace{\scriptstyle \left(p, \dots, p \right)}_{D \text{terms}}}(X) \right).
$$
The truncated shifted-signature coefficient vector of $X$ at order $D$ is then a vector of length 
$$
\widetilde{s}_p(D) = \sum_{d = 1}^D p^d = \frac{p^{D+1}-p}{p-1} \text{ when } p \ge 2 \text{ and } \widetilde{s}_p(D) = D \text{ when } p = 1.
$$
\end{definition}

\begin{lemma}
Let $\Psi : \mathcal{T} \rightarrow \mathcal{T}$ be a non-decreasing surjection and 
\begin{align*}
Y: \mathcal{T} & \longrightarrow \mathbb{R}^p \\
t & \longmapsto X({\Psi(t)}).
\end{align*}
Then $Sig(X) = Sig(Y)$ : the signature of $X$ is invariant by time reparameterization.
\end{lemma}

\begin{lemma}
Let 
\begin{align*}
Y: \mathcal{T} & \longrightarrow \mathbb{R}^p \\
t & \longmapsto X(t) + a, a \neq 0.
\end{align*}
Then $Sig(X) = Sig(Y)$ : the signature of $X$ is invariant by translation.
\end{lemma}

\begin{lemma} \label{lemma:avoidinvariance}
It is possible to circumvent the invariance by translation by adding an observation point that takes the value 0 at the beginning of $X$. Furthermore, it is possible to avoid the invariance by time reparameterization by considering the time-augmented function 
\begin{align*}
\widetilde{X}: \mathcal{T} & \longrightarrow \mathbb{R}^{p+1} \\
t & \longmapsto \left(X(t)^\top, t\right)^\top
\end{align*}
instead of $X$.
\end{lemma}

\begin{theorem}[Proposition 2 from \cite{fermanian2022functional}] \label{th:approx} \phantom{linebreak} \\
Let $f: \mathcal{D} \rightarrow \mathbb{R}$ be a continuous function where $\mathcal{D} \subset \mathcal{C}(\mathcal{T},\mathbb{R}^p)$ is a compact subset such that for any $X\in \mathcal{D}$, $X(0) = \bm{0}_p$. \\
For $X\in \mathcal{D}$ , let $\widetilde{X}$ be the time-augmented function of $X$. \\
Then, for every $\delta>0$, there exists $D^* \in \mathbb{N}, \alpha^* \in \mathbb{R}, \beta^*_{D^*} \in \mathbb{R}^{\widetilde{s}_{p+1}(D^*)}$, such that, for any $X \in \mathcal{D}$,
$$
\left|f(X)- \alpha^* - \langle\beta^*_{D^*}, \widetilde{S}^{D^*}(\widetilde{X})\rangle\right| \le \delta,
$$
where $\langle\cdot, \cdot\rangle$ denotes the Euclidean scalar product on $\mathbb{R}^{s_{p+1}(D^*)}$.
\end{theorem}

\section{Signature-based spatial regression for functional covariates} \label{sec:methodology}

In the following we denote 
$\mathcal{M}_{a \times b}(\mathbb{R})$ the set of real matrices of size $a \times b$, 
$\mathcal{B}_{\widetilde{s}_p(D), \alpha}$ the ball composed by the real vectors of size $\widetilde{s}_p(D)$ with a Euclidean norm less than $\alpha$, $\bm{0}_p$ the column vector consisting of $p$ times the value 0 and $\bm{1}_n$ the column vector consisting of $n$ times the value 1. \\

Let $X \in \mathcal{C}(\mathcal{T},\mathbb{R}^p)$. 
We assume that $X(0)=\bm{0}_p$ and we denote $\widetilde{X}$ as its time-augmented function. Next, Theorem \ref{th:approx} prompted \cite{Fre2025} to consider the following spatial autoregressive model 
\begin{equation} \label{eq:model}
Y_i = \rho^* \dsum_{j=1}^n w_{ij,n} Y_j + \alpha^* + \langle \beta^*_{D^*}, \widetilde{S}^{D^*}(\widetilde{X}_i) \rangle + \varepsilon_i,  
\end{equation}
where the spatial dependency between the $n$ spatial units $s_1, \dots, s_n$ is described by an $n \times n$ non-stochastic spatial weight matrix $W_n = (w_{ij,n})_{1 \le i,j \le n}$ that it is common (but not essential) to row-normalize in practice. The autoregressive parameter $\rho^*$ is in a compact space (typically $[-1,1]$), $\alpha^* \in \mathbb{R}$ and $\beta^*_{D^*} \in \mathbb{R}^{\widetilde{s}_{p+1}(D^*)}$. 
The disturbances $\{ \varepsilon_i, i = 1, \dots, n \}$ are assumed to be independent and identically distributed random variables independent of $\{  X_i(t), t \in \mathcal{T}, i = 1, \dots, n\}$, such that $\mathbb{E}(\varepsilon_i) = 0$ and $\mathbb{E}(\varepsilon_i^2) = \sigma^{2*}$. \\

However, estimating this model is complicated by two issues: (i) $D^*$ is unknown in practice and (ii) the size $\widetilde{s}_{p+1}(D^*)$ of $\beta^*_{D^*}$ increases exponentially with $D^*$ and polynomially with $p$; hence, the model is not directly solvable.

As mentioned above, \cites{Fre2025} PenSSAR approach (based on ridge regression) performed better than her PCA-ProjSSAR approach. As also mentioned above, the estimation of PenSSAR is iterative and may therefore require a long computation time. In this section, we describe the adaptation (NaivePenSSAR) of the approach developed by \cite{frevent2024multivariate} to the case of a univariate response variable. We then describe a new approach based on PLS projection.

\subsection{A naive signature-based spatial regression (NaivePenSSAR) model} \label{sec:modelnaive}

\subsubsection{The model}
When $Y$ is univariate, the signature-based model developed by \cite{frevent2024multivariate} becomes

\begin{equation} \label{eq:modelmultiadapted}
\bm{Y}_n = \rho^* W_n \bm{Y}_n + \alpha^* \bm{1}_n + \bm{\xi}_n^{D^*} \beta_{D^*}^* + \bm{\varepsilon}_n,    
\end{equation}
where $\bm{Y} = (Y_1, \dots, Y_n)^\top \in \mathbb{R}^n$,
$\bm{\xi}_n^{D^*} = (\widetilde{S}^{D^*}(\widetilde{X}_1)^\top, \dots, \widetilde{S}^{D^*}(\widetilde{X}_n)^\top)^\top \in \mathcal{M}_{n \times \widetilde{s}_{p+1}(D^*)}(\mathbb{R})$, $\alpha^* \in \mathbb{R}$, $\beta^*_{D^*} \in \mathbb{R}^{\widetilde{s}_{p+1}(D^*)}$, and $\bm{\varepsilon}_n = (\varepsilon_1, \dots, \varepsilon_n)^\top \in \mathbb{R}^n$.

Thus, \cites{frevent2024multivariate} model is equivalent to Model (\ref{eq:model}).

\subsubsection{Estimation}

The parameters $\rho^*, \alpha^*, \beta_{D^*}^*$ must be estimated. The true truncation order $D^*$ is also unknown and must be selected appropriately. Due to the large size of $\beta_{D^*}^*$, \cite{frevent2024multivariate} used a penalized ridge regression similar to the PenSSAR approach developed by \cite{Fre2025}. In contrast to  \cite{Fre2025}, however, \cites{frevent2024multivariate} estimation algorithm was not iterative and thus usually required shorter computation times.

We assume that $(\mathcal{H}_\alpha): \exists \alpha >0 / \beta_{D^*}^* \in \mathcal{B}_{\widetilde{s}_{p+1}(D^*),\alpha} $. \\

Next, for a fixed truncation order $D$, Model (\ref{eq:modelmultiadapted}) can be rewritten as 
$$ \bm{Y}_n = \rho_D^* W_n \bm{Y}_n + \alpha_D^* \bm{1}_n + \bm{\xi}_n^{D} \beta_{D}^* + \bm{\varepsilon}_n,$$ and, in a univariate context, the empirical objective function considered by \cite{frevent2024multivariate} becomes
$$ \widehat{\mathcal{R}}_D(\alpha_D,\beta_D,\rho_D) = \dfrac{1}{n} \left|\left| \bm{Y}_n - \rho_D W_n \bm{Y}_n - \alpha_D \bm{1}_n - \bm{\xi}_n^{D} \beta_D \right|\right|^2 $$
for $\alpha_D \in \mathbb{R}, \beta_D \in \mathcal{B}_{\widetilde{s}_{p+1}(D),\alpha}, \rho_D \in [-1,1]$. \\

This is equivalent to minimizing
$$ \widehat{\mathcal{R}}_D(\alpha_D,\beta_D,\rho_D) + \lambda ||\beta_D||^2 = \dfrac{1}{n} \left|\left| \bm{Y}_n - \rho_D W_n \bm{Y}_n - \alpha_D \bm{1}_n - \bm{\xi}_n^{D} \beta_D \right|\right|^2 + \lambda ||\beta_D||^2 $$
on $\mathbb{R} \times \mathbb{R}^{\widetilde{s}_{p+1}(D)} \times [-1,1]$,
where the ridge parameter $\lambda$ depends on $\alpha$. \\

By denoting
$$ (\widehat{\alpha}_D, \widehat{\beta}_D, \widehat{\rho}_D) = \underset{\substack{\alpha_D \in \mathbb{R} \\ \beta_D \in \mathbb{R}^{\widetilde{s}_{p+1}(D)} \\ \rho_D \in [-1,1]}}{\arg \min} \ \dfrac{1}{n} \left|\left| \bm{Y}_n - \rho_D W_n \bm{Y}_n - \alpha_D \bm{1}_n - \bm{\xi}_n^{D} \beta_D \right|\right|^2 + \lambda ||\beta_D||^2, $$

$\widehat{\alpha}_D, \widehat{\beta}_D$ have explicit formulas that depend on $\rho_D$:
$$ \begin{pmatrix}
\widehat{\alpha}_D(\rho_D) \\ \widehat{\beta}_D(\rho_D)
\end{pmatrix} = (\bm{\xi}_n^{'D\top} \bm{\xi}_n^{'D} + n \Lambda)^{-1} \bm{\xi}_n^{'D\top} (\bm{Y}_n - \rho_D W_n \bm{Y}_n), $$
where $\bm{\xi}_n^{'D} = (\bm{1}_n, \bm{\xi}_n^D)$ and $\Lambda = \begin{pmatrix}
    0 & 0 & 0 & \dots & 0 \\
    0 & \lambda & 0 & \dots & 0 \\
    0 & 0 & \ddots & \ddots & \vdots \\
    \vdots & \ddots & \ddots & \ddots & 0 \\
    0 & \dots & 0 & 0 & \lambda
\end{pmatrix}$.

Then, Algorithm \ref{algo:naive} allows one to compute $\widehat{\alpha}_D$, $\widehat{\beta}_D$, and $\widehat{\rho}_D$ for a fixed truncation order $D$ and a fixed regularization parameter $\lambda$. Consideration of this algorithm prompts two remarks:
\begin{remark}
In practice, the true truncated order $D^*$ is unknown. An estimator of $D^*$ can be defined in several ways (see \cite{fermanian2022functional,frevent2024multivariate}). In the following, we optimized $D^*$ on a validation set - as did \cite{Fre2025}.
\end{remark}

\begin{remark}
The objective function can be considered to be naive because it ignores the potential endogeneity issue between $W_n \bm{Y}_n$ and $\bm{\varepsilon}_n$. That is why we called this method ``NaivePenSSAR''. However, \cite{ma2020naive} proved that the estimators obtained were consistent, under reasonable assumptions.
\end{remark}

\begin{footnotesize}
\SetKwComment{Comment}{/* }{ */}
\begin{algorithm}
\caption{An algorithm for estimating the NaivePenSSAR for a truncation order $D$}
\label{algo:naive}
\KwData{$W_n, \bm{Y}_n, \bm{\xi}_n^{'D}, \lambda$}
\KwResult{$\widehat{\alpha}_D, \widehat{\beta}_D, \widehat{\rho}_D$} \vspace{0.3cm}
$\Lambda = \begin{pmatrix}
    0 & 0 & 0 & \dots & 0 \\
    0 & \lambda & 0 & \dots & 0 \\
    0 & 0 & \ddots & \ddots & \vdots \\
    \vdots & \ddots & \ddots & \ddots & 0 \\
    0 & \dots & 0 & 0 & \lambda
\end{pmatrix}$ \; \vspace{0.2cm}
$\begin{aligned}
\widehat{\rho}_D 
&= \underset{\rho_D \in [-1,1]}{\arg \min} \ \dfrac{1}{n} \left|\left| \bm{Y}_n - \rho_D W_n \bm{Y}_n - \bm{\xi}_n^{'D} \left( \bm{\xi}_n^{'D\top} \bm{\xi}_n^{'D} + n \Lambda \right)^{-1} \bm{\xi}_n^{'D\top} (\bm{Y}_n - \rho_D W_n \bm{Y}_n) \right|\right|^2
\end{aligned}$ \Comment*[r]{Estimation of $\rho_D^*$} \vspace{0.2cm}
$\begin{pmatrix} \widehat{\alpha}_D \\ \widehat{\beta}_D \end{pmatrix} = \left( \bm{\xi}_n^{'D\top} \bm{\xi}_n^{'D} + n \Lambda \right)^{-1} \bm{\xi}_n^{'D\top} (\bm{Y}_n - \widehat{\rho}_D W_n \bm{Y}_n) $ \Comment*[r]{Estimation of $\alpha_D^*$ and $\beta_D^*$}
\end{algorithm}
\end{footnotesize}

\subsection{A new approach based on partial least squares regression: PLS-ProjSSAR}

\subsubsection{The model}

Here we consider the projection approach developed by \cite{Fre2025}. In contrast to \cites{Fre2025} focus on PCA-based projection, we focussed on PLS projection. Model (\ref{eq:model}) thus becomes
\begin{equation} \label{eq:pls}
    \bm{Y}_n \approx \rho^* W_n \bm{Y}_n + \alpha^* \bm{1}_n + \bm{\zeta}_n^{D^*,J_n^{D^*}} \Phi_{D^*,J_n^{D^*}}^* 
    + \bm{\varepsilon}_n,
\end{equation}
where $\bm{\zeta}_n^{D^*,J_n^{D^*}} = (\bm{\zeta}_n^{(1)D^*}, \dots, \bm{\zeta}_n^{(J_n^{D^*})D^*})$ and $\bm{\zeta}_n^{(1)D^*}, \dots, \bm{\zeta}_n^{(J_n^{D^*})D^*}$ are the $J_n^{D^*}$ first PLS scores associated with $\bm{Y}_n$ and $\bm{\xi}_n^{D^*}$, and $\Phi_{D^*, J_n^{D^*}}^* 
\in \mathbb{R}^{J_n^{D^*}}$.

\subsubsection{Estimation}

As the true truncation order $D^*$ and the optimal number of PLS scores $J_n^{D^*}$ are unknown, we rewrote Model (\ref{eq:pls}) for a given truncated order $D$ and a number $J_n^D$ of PLS scores:
\begin{equation} \label{eq:plsD}
    \bm{Y}_n \approx \rho_D^* W_n \bm{Y}_n + \alpha_D^* \bm{1}_n + \bm{\zeta}_n^{D,J_n^D} \Phi_{D,J_n^D}^* + \bm{\varepsilon}_n,
\end{equation}
where $\bm{\zeta}_n^{D,J_n^D} = (\bm{\zeta}_n^{(1)D}, \dots, \bm{\zeta}_n^{(J_n^D)D})$ and $\bm{\zeta}_n^{(1)D}, \dots, \bm{\zeta}_n^{(J_n^D)D}$ are the $J_n^D$ first PLS scores associated with $\bm{Y}_n$ and $\bm{\xi}_n^D$, $\mathbb{E}(\bm{\varepsilon}_n) = \bm{0}_n$ and $\mathbb{V}(\bm{\varepsilon}_n) = \sigma_D^{2} I_n$. \\

Next, adopting \cites{huang2021robust} approach, we neglect the autoregressive term $\rho^*_D W_n \bm{Y}_n$ when computing these values. The subsequent calculation can then be carried out using Algorithm \ref{algo:PLSreg}.

\begin{footnotesize}
\SetKwComment{Comment}{/* }{ */}
\begin{algorithm}
\caption{An algorithm for estimating the PLS scores $\bm{\zeta}_n^D$ for a truncation order $D$ and a number $J_n^D$ of PLS scores}
\label{algo:PLSreg}
\KwData{$\bm{Y}_n, \bm{\xi}_n^D, J_n^D$}
\KwResult{$\bm{\zeta}_n^{D,J_n^D} = 
(\bm{\zeta}_n^{(1)D}, \dots, \bm{\zeta}_n^{(J_n^D)D})$} \vspace{0.3cm}
$\bm{\xi}_n^{(0)D} = \bm{\xi}_n^D$ ; $\bm{Y}_n^{(0)D} = \bm{Y}_n$ \; \vspace{0.2cm}
\For{$k = 1$ \KwTo $J_n^D$}{ 
\vspace{0.2cm}
    $w^{(k)D} = \dfrac{\bm{\xi}_n^{(k-1)D\top} \bm{Y}_n^{(k-1)} }{|| \bm{\xi}_n^{(k-1)D\top} \bm{Y}_n^{(k-1)} ||}$ \; \vspace{0.3cm}
    $\bm{\zeta}_n^{(k)D} = \bm{\xi}_n^{(k-1)D} w^{(k)D}$ \; \vspace{0.3cm}
    $p^{(k)D} = \dfrac{\bm{\xi}_n^{(k-1)D\top} \bm{\zeta}_n^{(k)D}}{\bm{\zeta}_n^{(k)D\top} \bm{\zeta}_n^{(k)D}} $ \; \vspace{0.3cm}
    $q^{(k)D} = \dfrac{\bm{Y}_n^{(k-1)D\top} \bm{\zeta}_n^{(k)D}}{\bm{\zeta}_n^{(k)D\top} \bm{\zeta}_n^{(k)D}} $ \; \vspace{0.3cm}
    $\bm{\xi}_n^{(k)D} = \bm{\xi}_n^{(k-1)D} - \bm{\zeta}_n^{(k)D} p^{(k)D\top}$ \;
    $\bm{Y}_n^{(k)D} = \bm{Y}_n^{(k-1)} - \bm{\zeta}_n^{(k)D} q^{(k)D\top}$ \;
}
\end{algorithm}
\end{footnotesize}

We then consider the quasi-log-likelihood associated with Model (\ref{eq:plsD}):
\begin{align} \label{eq:likelihoodPLS}
\widetilde{\ell}_n(\sigma^2_D,\rho_D,\Phi_{D,J_n^D}) &= - \dfrac{n}{2} \ln{(2\pi)} - \dfrac{n}{2} \ln{(\sigma^2_D)} + \ln{|I_n - \rho_D W_n|} \\ &- \dfrac{1}{2\sigma^2_D} \left[ (I_n - \rho_D W_n) \bm{Y}_n - \alpha_D \bm{1}_n - \bm{\zeta}_n^{D,J_n^D} \Phi_{D,J_n^D} \right]^\top \left[ (I_n - \rho_D W_n) \bm{Y}_n - \alpha_D \bm{1}_n - \bm{\zeta}_n^{D,J_n^D} \Phi_{D,J_n^D} \right]. \nonumber
\end{align}

For a fixed $\rho_D$, (\ref{eq:likelihoodPLS}) is maximized in
$$
(\widehat{\alpha}_D(\rho_D), \widehat{\Phi}_{D,J_n^D}(\rho_D)^\top)^\top = ( \bm{\zeta}_n^{'D,J_n^D \top} \bm{\zeta}_n^{'D,J_n^D} )^{-1} \bm{\zeta}_n^{'D,J_n^D \top} (I_n - \rho_D W_n) \bm{Y}_n 
$$
and
$$
\widehat{\sigma^2_D}(\rho_D) = \dfrac{1}{n} \left[ (I_n - \rho_D W_n) \bm{Y}_n - \bm{\zeta}_n^{'D,J_n^D} (\widehat{\alpha}_D, \widehat{\Phi}_{D,J_n^D}^\top)^\top \right]^\top \left[ (I_n - \rho_D W_n) \bm{Y}_n - \bm{\zeta}_n^{'D,J_n^D} (\widehat{\alpha}_D, \widehat{\Phi}_{D,J_n^D}^\top)^\top \right],
$$
where $\bm{\zeta}_n^{'D,J_n^D} = (\bm{1}_n, \bm{\zeta}_n^{D,J_n^D})$.

The concentrated quasi-log-likelihood is then defined as
\begin{equation*}
\widetilde{\ell}_n(\rho_D) = - \dfrac{n}{2} \left[ \ln{(2\pi)} + 1 \right] - \dfrac{n}{2} \ln{(\widehat{\sigma^2_D}(\rho_D))} + \ln{|I_n - \rho_D W_n|}.
\end{equation*}

Next, $\rho_D^*$ is estimated by $\widehat{\rho}_D = \underset{\rho_D \in [-1,1]}{\arg\max} \ \widetilde{\ell}_n(\rho_D)$, and $\sigma^{2*}_D$, $\alpha^*_D$ and $\Phi_{D,J_n^D}^*$ are estimated respectively by
$\widehat{\sigma^2_D}(\widehat{\rho}_D)$, $\widehat{\alpha}_D(\widehat{\rho}_D)$ and $ \widehat{\Phi}_{D,J_n^D}(\widehat{\rho}_D) $.

Consideration of the whole algorithm for the PLS-ProjSSAR (Algorithm \ref{algo:PLSSAR}) prompts two further remarks.

\begin{remark}
As with the NaivePenSSAR approach, $D^*$ can be selected from a predefined set of possible values $\{1, \dots, D^\text{max}\}$ in a validation set.\end{remark}

\begin{remark}
The optimal number of PLS components $J_n^D$ in Model (\ref{eq:plsD}) and Algorithm \ref{algo:PLSreg} can be selected in several ways, for example by cross-validation for each considered truncation order $D \in \{1, \dots, D^\text{max}\}$ before selecting the best number $\widehat{J}_n^{\widehat{D}}$ associated with the best truncation order $\widehat{D}$. Here, we select both values using the same validation set. We chose the truncation order $\widehat{D}$ and the corresponding optimal number of PLS components $\widehat{J}_n^{\widehat{D}}$ that gave the smallest error on the validation set.
\end{remark}

\begin{footnotesize}
\SetKwComment{Comment}{/* }{ */}
\begin{algorithm}
\caption{An algorithm for estimating the PLS-ProjSSAR model for a truncation order $D$ and a number of PLS scores $J_n^D$}
\label{algo:PLSSAR}
\KwData{$W_n, \bm{Y}_n, \bm{\xi}_n^D, J_n^D$}
\KwResult{$\widehat{\alpha}_D, \widehat{\Phi}_{D,J_n^D}, \widehat{\rho}_D, \widehat{\sigma^2_D}$} \vspace{0.3cm}
Computation of $\bm{\zeta}_n^{D,J_n^D} = (\bm{\zeta}_n^{(1)D}, \dots, \bm{\zeta}_n^{(J_n^D)D})$ using Algorithm \ref{algo:PLSreg} \;
$\widehat{\rho}_D = \underset{\rho_D \in [-1,1]}{\arg\max} \ -\dfrac{n}{2} \ln{\left[ \dfrac{1}{n} \left\| (I_n - \rho_D W_n) \bm{Y}_n - \bm{\zeta}_n^{'D,J_n^D} ( \bm{\zeta}_n^{'D,J_n^D \top} \bm{\zeta}_n^{'D,J_n^D} )^{-1} \bm{\zeta}_n^{'D,J_n^D \top} (I_n - \rho_D W_n) \bm{Y}_n \right\|^2 \right] } $ \\
\hspace{2.15cm} $+ \ln{|I_n - \rho_D W_n|}$ \Comment*[r]{Estimation of $\rho_D^*$}
$(\widehat{\alpha}_D, \widehat{\Phi}_{D,J_n^D}^\top)^\top = ( \bm{\zeta}_n^{'D,J_n^D \top} \bm{\zeta}_n^{'D,J_n^D} )^{-1} \bm{\zeta}_n^{'D,J_n^D \top} (I_n - \widehat{\rho}_D W_n) \bm{Y}_n $ \Comment*[r]{Estimation of $\alpha_D^*$ and $\Phi_{D,J_n^D}^*$}
$\widehat{\sigma^2_D} = \dfrac{1}{n} \left\| (I_n - \widehat{\rho}_D W_n) \bm{Y}_n - \widehat{\alpha}_D \bm{1}_n - \bm{\zeta}_n^{D,J_n^D} \widehat{\Phi}_{D,J_n^D} \right\|^2 $ \Comment*[r]{Estimation of $\sigma^{2*}_D$}
\end{algorithm}
\end{footnotesize}

\subsubsection{Linkage to ProjSSAR and theoretical guaranties}

When considering $D^* = D = +\infty$, Model (\ref{eq:model}) corresponds exactly to Model (7) from \cite{Fre2025} and Equation (\ref{eq:likelihoodPLS}) is the PLS version of Equation (9) from \cite{Fre2025}. \\
In this case, considering that the $\bm{\zeta}_n^{(k)D}$ are centered (this can be easily satisfied by subtracting their average) and that $\alpha^*_D = 0$ (which can be obtained by centering $\bm{Y}_n$) and by making the same assumptions as in the PCA-ProjSSAR developed by \cite{Fre2025}, the latter's Theorems 2.7, 2.8 and 2.9 are valid.

\section{The simulation study} \label{sec:simu}

In a simulation study with the same design as that described in \cite{Fre2025}, we compare the performances of NaivePenSSAR and PLS-ProjSSAR with those of the FSARLM \citep{ahmed2022quasi}, PCA-ProjSSAR \citep{Fre2025} and PenSSAR \citep{Fre2025}.

\subsection{Study design}

We considered a grid with $60 \times 60$ locations and to which we randomly allocated $n=200$ spatial units.

The data were generated using the following models: 
\begin{itemize}[leftmargin=1.9cm]
\item[Model 1.] $Y_i = \rho^* \dsum_{j=1}^n w_{ij,n} Y_j + \dfrac{1}{p} \dsum_{k=1}^p X_{i,k}(t_{101}) + \varepsilon_i $
\item[Model 2.] $Y_i = \rho^* \dsum_{j=1}^n w_{ij,n} Y_j + \dfrac{1}{p} \dsum_{k=1}^p Z_{i,k}(t_{101}) + \varepsilon_i $,
\end{itemize}

where $\varepsilon_i \sim \mathcal{N}(0,1)$, $X_i$ was generated in 101 equally spaced times of $[0,1]$ ($t_1, \dots, t_{101}$) as
$$ X_i(t) = (X_{i,1}(t), \dots, X_{i,p}(t))^\top, X_{i,k}(t) = \alpha_{i,k}t + f_{i,k}(t),$$
where $\alpha_{i,k} \sim \mathcal{U}([-3,3])$, $f_{i,k}$ is a Gaussian process with exponential covariance matrix with length-scale 1, and $Z_i$ was generated in $t_1, \dots, t_{101}$ as
$$ Z_i(t) = (Z_{i,1}(t), \dots, Z_{i,p}(t))^\top, Z_{i,k}(t) = \beta_{i,k,1}t + 10 \beta_{i,k,2} \sin{\left( \dfrac{2 \pi t}{\beta_{i,k,3}} \right)} + 10 (t - \beta_{i,k,4})^3, $$
where $\beta_{i,k,j} \sim \mathcal{U}([0,1]), j = 1,2,3,4$.

We sought to predict $Y_i$ from the 100 first observations of $X_i$ (for Model 1) and the 100 first observations of $Z_i$ (for Model 2).

We used the $k$-nearest neighbors method (with $k = 4$ or $8$) to build the spatial weight matrix $W_n$, and we considered $p = 2,6,10$ and $\rho^* = 0, 0.2, 0.4, 0.6, 0.8$.

For each model and each value of $k$, $p$ and $\rho^*$, several approaches were compared:
\begin{itemize}
\item[(i)] The FSARLM developed by \cite{ahmed2022quasi}, using a cubic B-spline basis with 12 equally spaced knots to approximate the $X_i$ ($Z_i$) from the observed data and a functional PCA \citep{ramsay2005functional}. We considered a maximum number of coefficients such that the cumulative inertia was below 95\% as suggested by \cite{ahmed2022quasi}.
\item[(ii)] The PenSSAR developed by \cite{Fre2025}, with a maximum truncation order such that there were at most $10^4$ signature coefficients.
\item[(iii)] The PCA-ProjSSAR developed by \cite{Fre2025}, with a maximum truncation order to give at most $10^4$ signature coefficients, a PCA on the standardized truncated signature coefficient vectors, and a maximum number of coefficients such that the cumulative inertia was below 95\%.
\item[(iv)] The NaivePenSSAR approach, with at most $10^4$ signature coefficients.
\item[(v)] The PLS-ProjSSAR approach, with at most $10^4$ signature coefficients.
\end{itemize}

To this end, 200 datasets were generated. Each was split into a training set, a validation set, and a test set such that the optimal number of coefficients
(for the FSARLM), the optimal truncation order (for PenSSAR and NaivePenSSAR), and the optimal number of coefficients associated with the optimal truncation order (for PCA-ProjSSAR and PLS-ProjSSAR) was selected in the validation set on the basis of the root mean squared error (RMSE) criterion. The level of performance of each approach was evaluated on the test set, using the RMSE. To generate the training, validation and test sets, we considered ordinary validation and spatial validation. For the latter,
we applied a $K$-means algorithm (with $K=6$) to the coordinates of the data; two clusters were selected at random to be the validation and test sets. \\

It should be noted that for the signature-based approaches, we avoided invariance by applying the strategy described in Lemma \ref{lemma:avoidinvariance}.

\subsection{The results of the simulation study}

The results for $k = 4$ are presented in Figures \ref{fig:model1vois4} and \ref{fig:model2vois4}, and the results for $k = 8$ are presented in Appendix \ref{appendix:ressim} (Figures \ref{fig:model1vois8} and \ref{fig:model2vois8}). 

The signature-based approaches performed at least as well as the FSARLM; this was particularly this case for Model 2, in which the signature-based approaches presented much lower RMSEs. These results are in line with those reported by \cite{Fre2025}. \\
The NaivePenSSAR and PLS-ProjSSAR's levels of performance were very similar to that of PenSSAR. With both models, these three approaches almost always gave the best results. However, NaivePenSSAR and especially PLS-ProjSSAR presented much shorter computation times (Table \ref{tab:time}), which should facilitate the application of these two methods.

\begin{figure}
\centering
\includegraphics[width=\textwidth]{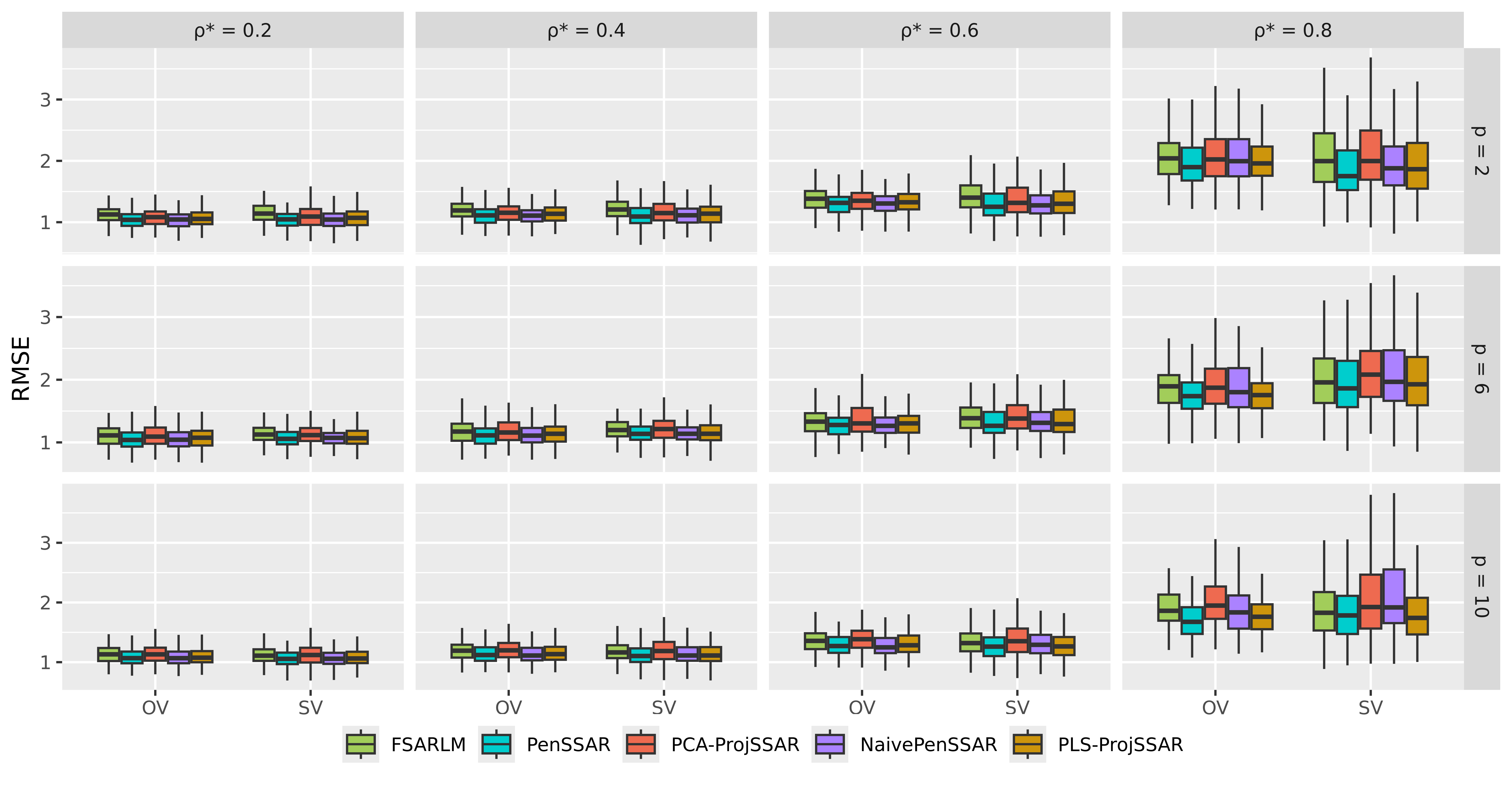}
\caption{RMSEs for Model 1 on the test set with the FSARLM, PenSSAR, PCA-ProjSSAR, NaivePenSSAR and PLS-ProjSSAR methods, using ordinary validation (OV) and spatial validation (SV), and with $k = 4$.}
\label{fig:model1vois4}
\end{figure}

\begin{figure}
\centering
\includegraphics[width=\textwidth]{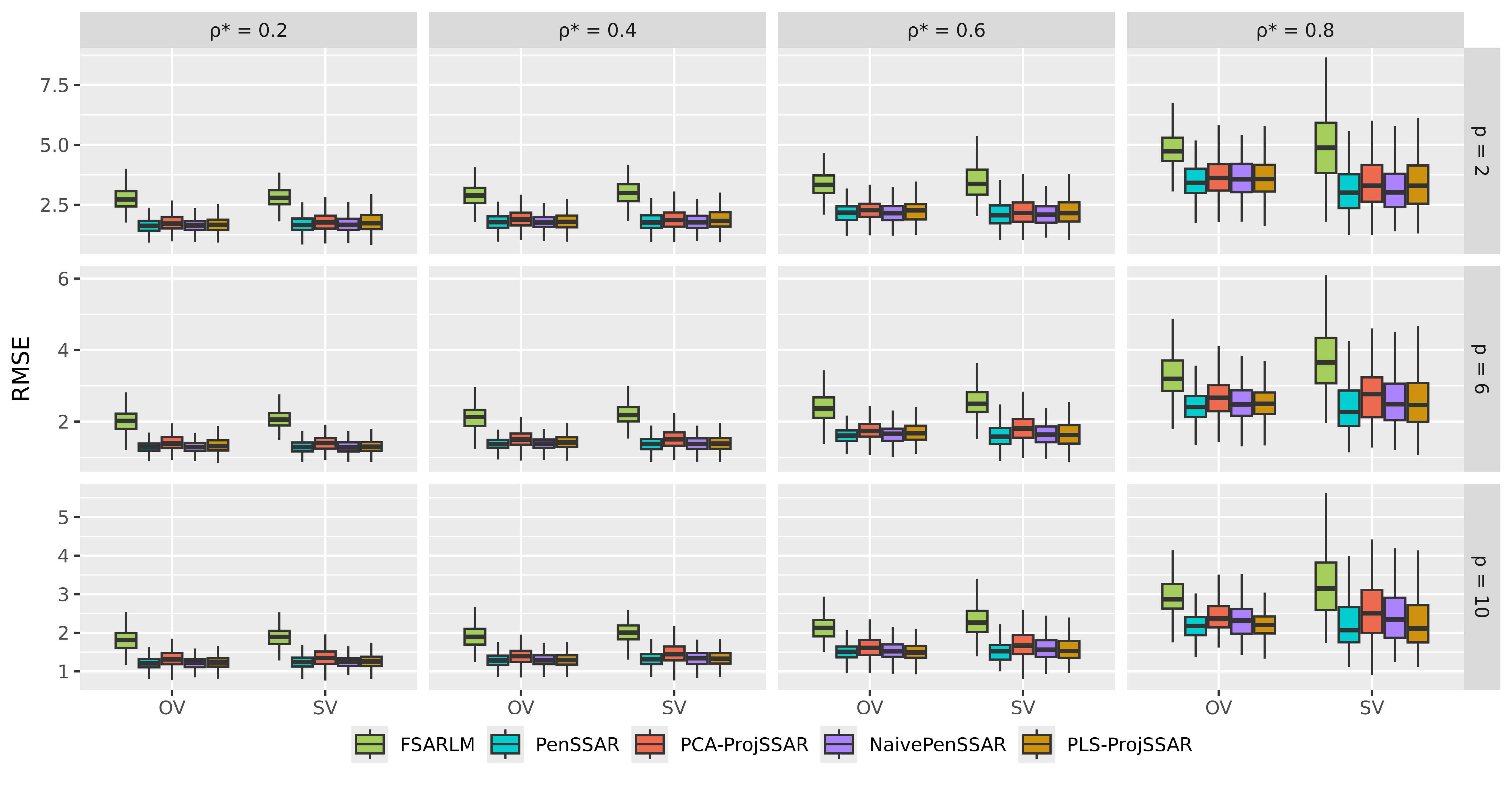}
\caption{RMSEs for Model 2 on the test set with the FSARLM, PenSSAR, PCA-ProjSSAR, NaivePenSSAR and PLS-ProjSSAR methods, using ordinary validation (OV) and spatial validation (SV), and with $k = 4$.}
\label{fig:model2vois4}
\end{figure}

\begin{table}[]
\caption{The median [interquartile range] computation time (in seconds) for the FSARLM, PenSSAR, PCA-ProjSSAR, NaivePenSSAR and PLS-ProjSSAR methods in the simulation study.}
\label{tab:time}
\centering
\begin{tabular}{lccc}
\hline
 & \multicolumn{3}{c}{\textbf{Model 1}} \\ \hline
 & $\mathbf{p=2}$    & $\mathbf{p=6}$   & $\mathbf{p=10}$  \\ 
\textbf{FSARLM} & 0.53 [0.46, 0.60] & 1.15 [1.10, 1.33] & 2.32 [2.25, 2.39] \\
\textbf{PenSSAR} & 1857.08 [1361.96,2687.22] & 58.84 [38.14, 86.47] & 4.61 [3.82, 6.35] \\
\textbf{PCA-ProjSSAR} & 10.02 [8.84, 11.74] & 7.16 [6.41, 8.01] & 4.49 [4.17, 4.86] \\
\textbf{NaivePenSSAR} & 1465.95 [1246.10,1819.63] & 73.56 [53.48, 91.95] & 5.18 [4.62, 6.35] \\
\textbf{PLS-ProjSSAR} & 22.56 [20.01,25.55] & 9.57 [8.02, 11.58] & 10.79 [8.80, 13.10] \\ \hline \hline
 & \multicolumn{3}{c}{\textbf{Model 2}} \\ \hline
 & $\mathbf{p=2}$    & $\mathbf{p=6}$   & $\mathbf{p=10}$   \\ 
\textbf{FSARLM} & 0.80 [0.66, 0.91] & 1.75 [1.64, 1.99] & 3.20 [3.09, 3.34] \\
\textbf{PenSSAR} & 30171.64 [17964.63, 49477.27] & 80.65 [49.89, 142.81] & 11.18 [8.33, 16.16] \\
\textbf{PCA-ProjSSAR} & 8.20 [6.88, 9.70] & 5.89 [5.39, 6.44] & 3.73 [3.48, 4.02] \\
\textbf{NaivePenSSAR} & 1537.43 [1237.53, 1936.07] & 68.24 [48.81, 84.95] & 5.12 [4.46, 5.92] \\
\textbf{PLS-ProjSSAR} & 22.25 [19.01, 26.16] & 9.28 [7.71, 10.83] & 5.67 [5.09, 6.11] \\ 
\hline
\end{tabular}
\end{table}

\section{Application to a real dataset} \label{sec:appli}

\begin{figure}
\centering
\includegraphics[width=0.49\linewidth]{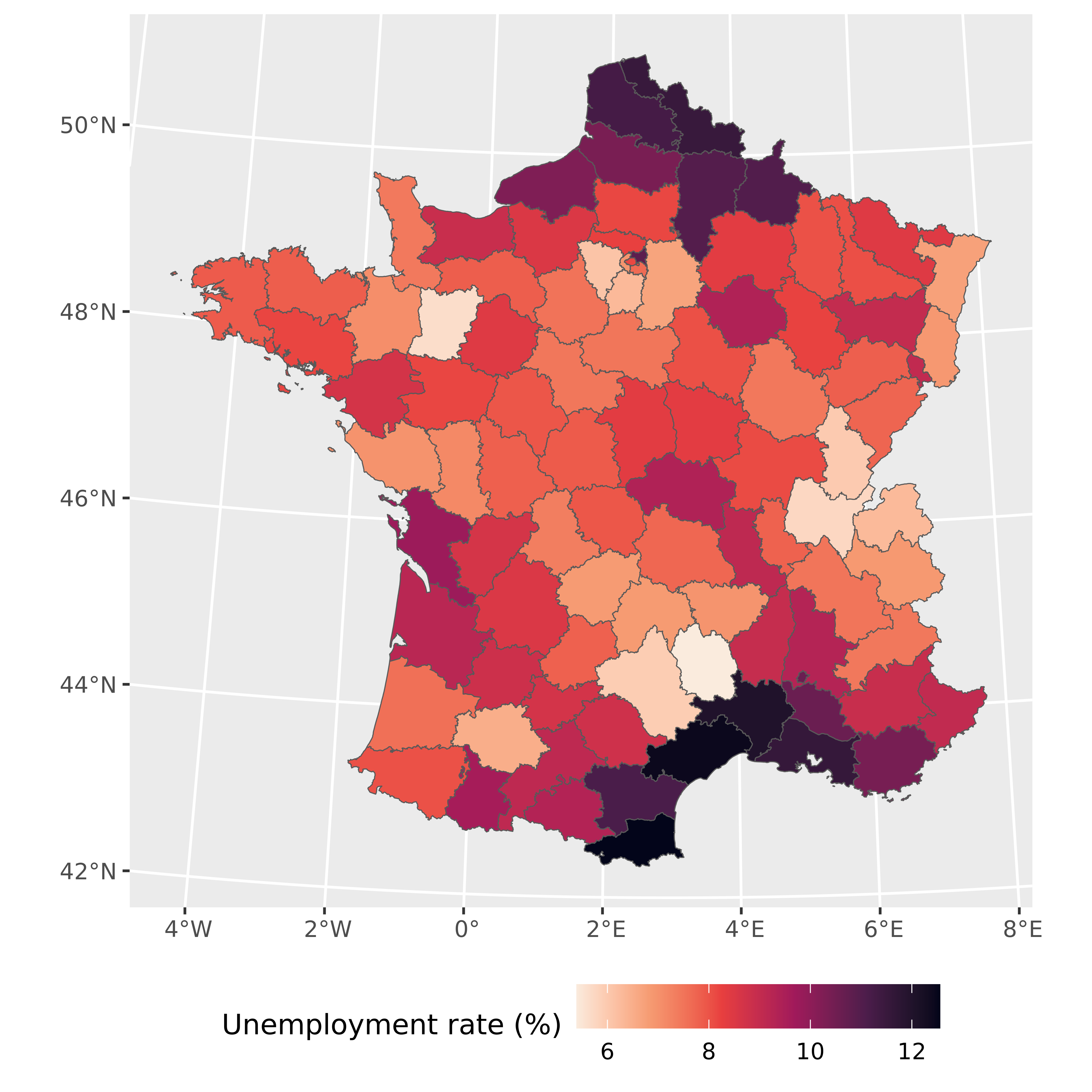}
\includegraphics[width=0.49\linewidth]{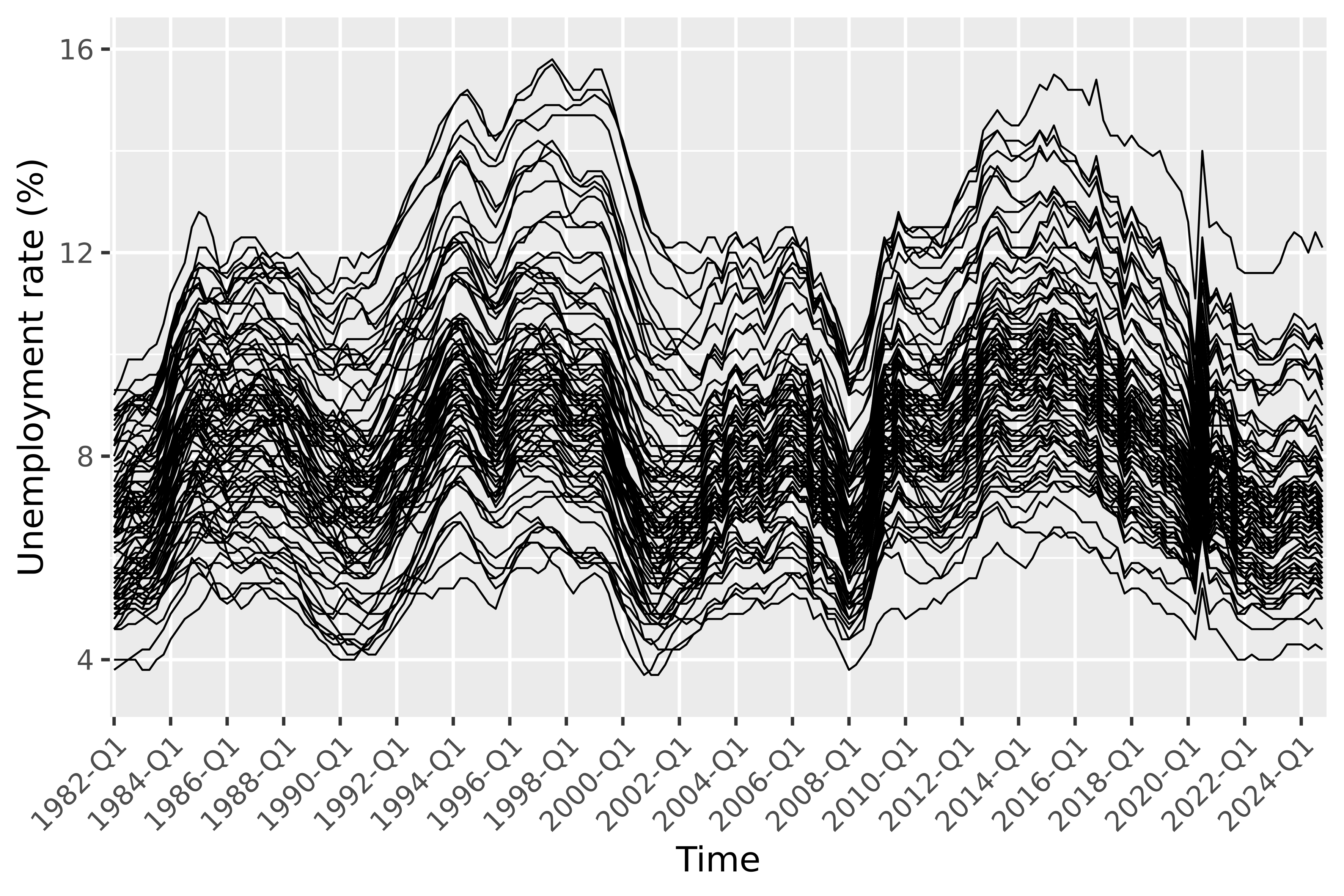}
\caption{Spatial distribution of the mean unemployment rate over the period 1982-2024 in each of the 94 \textit{départements} in mainland France (left panel), and the changes in the \textit{départements}' unemployment rates over the same period (right panel).}
\label{fig:chomage}
\end{figure}

\begin{figure}
\centering
\includegraphics[width=0.49\linewidth]{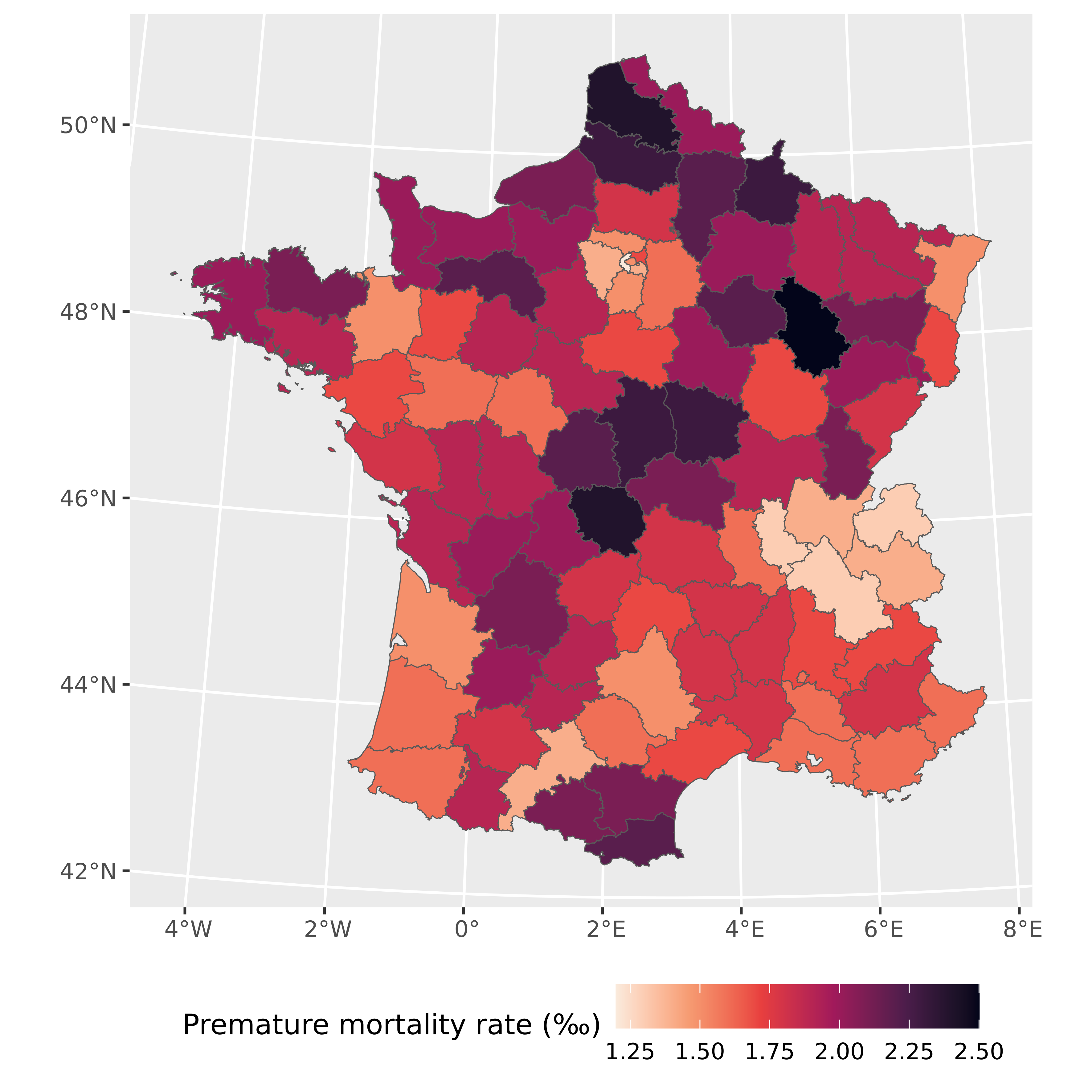}
\includegraphics[width=0.49\linewidth]{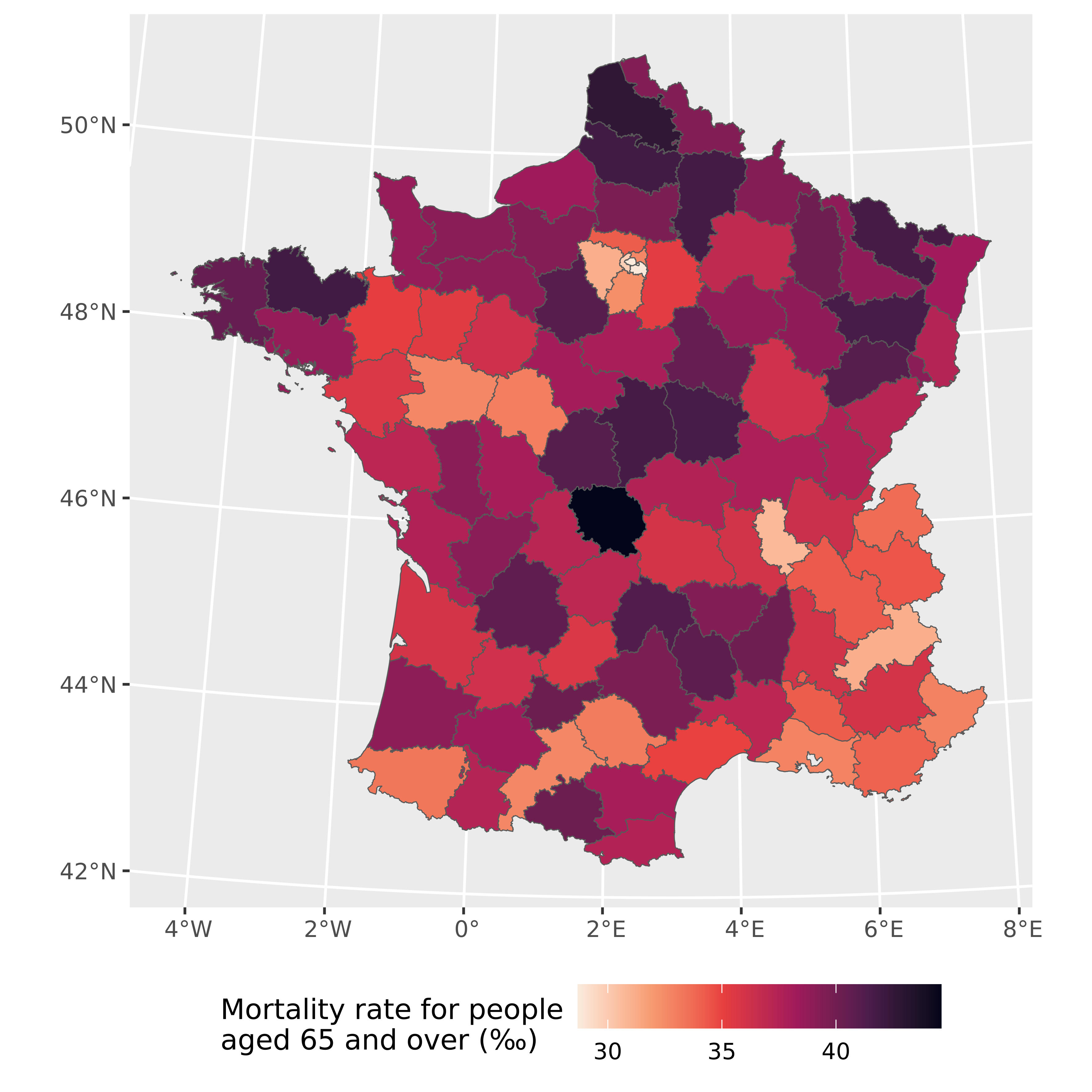}
\caption{Spatial distribution of the premature mortality rate in 2024 (left panel) and the mortality rate among people aged 65 and over in 2024 (right panel), in each of the 94 \textit{départements} in mainland France.}
\label{fig:death}
\end{figure}

We analyzed data on the premature mortality rate (per 1000 people) and the mortality rate (per 1000 people aged 65 and over) in the 94 administrative counties (\textit{départements}) of mainland France in 2024 (provided by the \textit{Institut National de la Statistique et des Etudes Economiques} (Paris, France)). The objective was to predict these mortality rates from the unemployment rates in each quarter over the period 1982-2024 (Figures \ref{fig:chomage} and \ref{fig:death}).

We use the FSARLM, PenSSAR, PCA-ProjSSAR, NaivePenSSAR and PLS-ProjSSAR methods, considering spatial weight matrices based on the $k$-nearest neighbors (with $k=4$ or $8$). 

We also considered ordinary validation and spatial validation by applying a $K$-means algorithm (with $K$ = 6) to the coordinates of the data. To covering all the possibilities for spatial validation, we repeated the procedure on 30 different train/validation/test sets.

\begin{figure}
\centering
\includegraphics[width=0.8\linewidth]{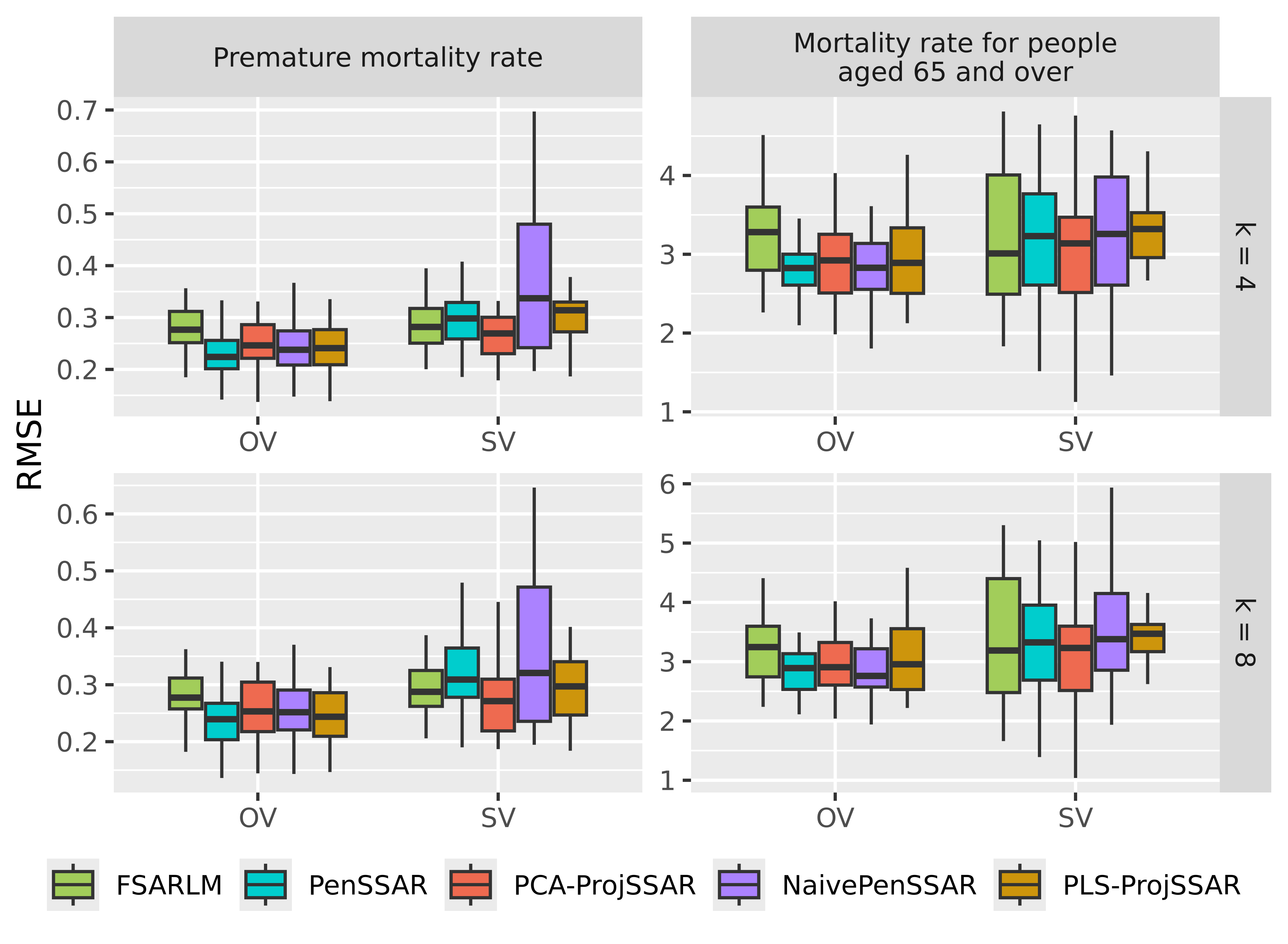}
\caption{RMSEs from 30 test sets for predicting the premature mortality rate and the mortality rate among people aged 65 and over with the FSARLM, PenSSAR, PCA-ProjSSAR, NaivePenSSAR and PLS-ProjSSAR methods.}
\label{fig:resappli}
\end{figure}

With both datasets, the signature approaches present competitive performances compared with the FSARLM, especially when using an ordinary validation (Figure \ref{fig:resappli}). PLS-ProjSSAR performed similarly to PenSSAR but presenting much shorter computation times (Table \ref{tab:timeappli}).

\begin{table}[h]
\caption{The median [interquartile range] computation time (in seconds) for the FSARLM, PenSSAR, PCA-ProjSSAR, NaivePenSSAR and PLS-ProjSSAR in the prediction of the premature mortality rate and the mortality rate for people aged 65 and over in 2024.}
\label{tab:timeappli}
\centering
\begin{tabular}{lcc}
\hline
 & \textbf{Premature mortality rate} & \textbf{Mortality rate for people aged 65 and over} \\ \hline
\textbf{FSARLM} & 0.61 [0.57, 0.73] & 0.63 [0.58, 0.64] \\
\textbf{PenSSAR} & 139585.44 [114924.69, 172469.80] & 411207.98 [277403.83, 503575.54] \\
\textbf{PCA-ProjSSAR} & 2.18 [1.92, 3.09] & 2.09 [1.79, 2.93] \\
\textbf{NaivePenSSAR} & 736.08 [714.32, 787.09] & 1103.73 [773.60, 1304.68] \\
\textbf{PLS-ProjSSAR} & 6.48 [5.73, 7.17] & 9.30 [6.59,10.91] \\ 
\hline
\end{tabular}
\end{table}

\section{Discussion} \label{sec:discussion}

Here, we developed two new alternatives to signature-based, spatial autoregressive models \citep{Fre2025}. The first (NaivePenSSAR) is an adaptation of the approach developed by \cite{frevent2024multivariate} and was applied to the case of a univariate response variable. The second (PLS-ProjSSAR) considered signature projections through the application of a PLS algorithm. \\

The results of our simulation study showed that the signature-based models performed better than a conventional model that does not use signatures \citep{ahmed2022quasi}. Furthermore, the new models had a level of performance that was similar to that of the signature-based PenSSAR approach \citep{Fre2025}. However, the new models presented much shorter computation times; this should facilitate their application. The same conclusion was drawn from application of the new models (especially PLS-ProjSSAR) to the premature mortality rate and the mortality rate for people aged 65 and over. \\

In the present work, we focused on functional covariates. However, it should be borne in mind that both functional and non-functional covariates can be integrated into our models. Furthermore, both of the new methods can be adapted for use with other spatial models, such as spatial error models. \\

Lastly, it should be noted that just as NaivePenSSAR is the univariate version of the model developed by \cite{frevent2024multivariate}, PLS-ProjSSAR could also be adapted for use with a multivariate response variable; we intend to address this aspect in the future.

\bibliography{bibliographie.bib}
\bibliographystyle{chicago}

\appendix

\section{Supplementary materials for the simulation study ($k = 8$)} \label{appendix:ressim}

\begin{figure}
\centering
\includegraphics[width=\textwidth]{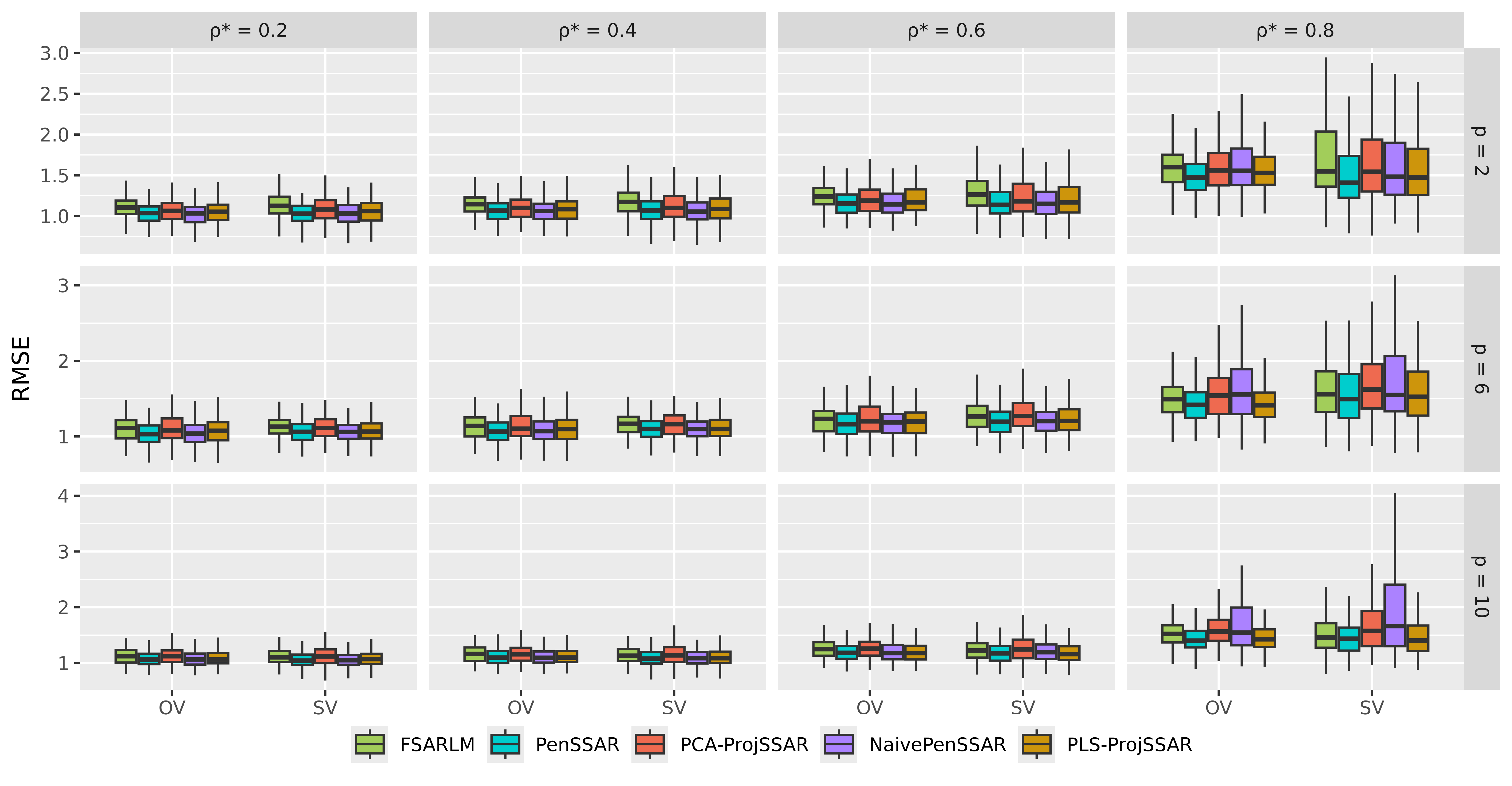}
\caption{RMSEs for Model 1 on the test set with the FSARLM, PenSSAR, PCA-ProjSSAR, NaivePenSSAR and PLS-ProjSSAR methods, using ordinary validation (OV) and spatial validation (SV), and with $k = 8$.}
\label{fig:model1vois8}
\end{figure}

\begin{figure}
\centering
\includegraphics[width=\textwidth]{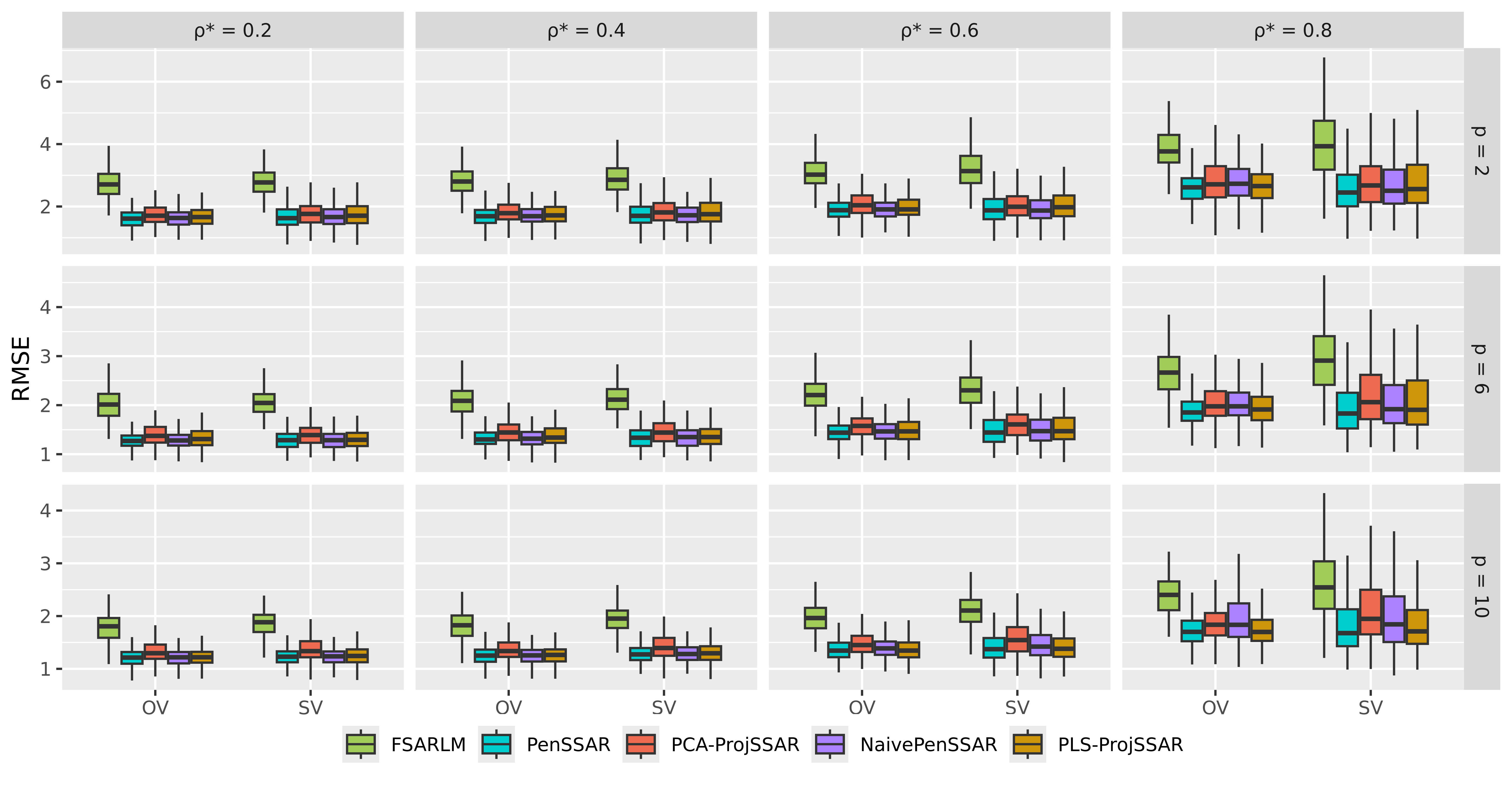}
\caption{RMSEs for Model 2 on the test set with the FSARLM, PenSSAR, PCA-ProjSSAR, NaivePenSSAR and PLS-ProjSSAR methods, using ordinary validation (OV) and spatial validation (SV), and with $k = 8$.}
\label{fig:model2vois8}
\end{figure}

Figure \ref{fig:model1vois8} and \ref{fig:model2vois8} show the RMSEs computed for the test set, using ordinary validation (OV) and spatial validation (SV).

\end{document}